# Quantifying sulfur speciation in magmatic-hydrothermal fluids

**Gleb S. Pokrovski**

*Experimental Geosciences Team (GeoExp), Géosciences Environnement Toulouse (GET), CNRS, University of Toulouse, IRD, CNES, 14 avenue Edouard Belin, 31400 Toulouse France*

gleb.pokrovski@get.omp.eu; glebounet@gmail.com

phones: +33 5 61 33 26 18; +33 6 20 34 72 62

Quantitative knowledge of sulfur speciation in the fluid phase is key to understanding sulfur degassing from magmas and its transfer along with metals by fluids across the lithosphere. Farsang and Zajacz[1] recently reported new sulfur speciation data using Raman spectroscopy measurements on aqueous $H_2SO_4$-NaCl(-KCl) solutions trapped as synthetic fluid inclusions in quartz at 875 °C and 2 kbar under variable redox conditions. They interpreted the data by dominant $SO_{2(aq)}$ and $HS^-$, along with subordinate $H_2S_{(aq)}$, whereas the di- and trisulfur radical ions, $[S_3^{\bullet}]^-$ and $[S_2^{\bullet}]^-$, reported in aqueous fluids both in nature and experiment to at least 700 °C and 15 kbar[2-5], were undetectable in Raman non-resonant spectra[1]. On the basis of thermochemical calculations that returned negligible $HS^-$ and large $[S_3^{\bullet}]^-$ concentrations at their experimental conditions, the authors claimed that the published $[S_3^{\bullet}]^-$ and $HS^-$ thermodynamic data were incorrect and that the major species controlling both sulfur and gold transport in hydrothermal-magmatic fluids is the $HS^-$ anion. Here I demonstrate that their conclusions stem from the use of inconsistent thermodynamic data and incorrect Raman spectra assignments. Thus, ref.[1] provides no grounds for questioning the validity of the thermodynamic properties of $HS^-$ and $[S_3^{\bullet}]^-$ or of their metal complexes in the fluids of the Earth's crust.

The report by Farsang and Zajacz[1] of $HS^-$ fractions exceeding $H_2S_{(aq)}$ in their Raman spectra stands in direct contradiction with their adopted thermochemical model predicting negligible $HS^-$ and abundant $[S_3^{\bullet}]^-$. To enable unbiased comparisons, I reproduced their calculations in Fig. 1a (in $log_{10}$ scale enabling visualization of minor sulfur species), using the same thermodynamic data sources and software[1]. Surprisingly, the authors exploit this model to claim that predicted $[S_3^{\bullet}]^-$ concentrations were overestimated. If $HS^-$ were as abundant as suggested[1], the shown species distribution *a priori* cannot be used to support these claims. To match the $HS^-/H_2S$ ratios derived from Raman spectra at sulfide-dominated conditions (~2:1, Table S2)[1], the equilibrium constant value ($K_1$) of the reaction

$H_2S_{(aq)} = HS^- + H^+$ (1)

must be increased by at least 8 log units. The impact of this correction on S speciation is remarkable (Fig. 1b). First, $[S_3^{\bullet}]^-$ concentrations are ~4 log units lower than in Fig. 1a and would have been undetectable by Raman even in resonance regime (i.e. with 532 nm laser). Second, the apparently constant $HS^-/H_2S$ ratio reported over a wide $f_{O_2}$ range (Table S2; NNO–0.9 to NNO+2.6)[1] cannot be matched with a single value of $K_1$, implying that such an $HS^-$-dominated model would be wrong. Third, such anomalous dissociation of $H_2S_{(aq)}$ is at odds with our current knowledge of the thermochemistry of similar weak acids (Fig. S1). Fourth, if the $HS^-$ abundance in ref.[1] were correct, it would result in an unphysical gold speciation, with the $Au^+$ cation being dominant as imposed by electrical charge balance constraints to compensate for negatively charged $HS^-$ rather than increasing $AuHS_{(aq)}$ and $Au(HS)_2^-$ concentrations as the authors expected. Finally, the Raman feature around 2570 cm$^{-1}$, attributed to the stretching vibration of the $HS^-$ ion[1], rather belongs to the asymmetric $H_2S$ $\nu_1$ peak as evidenced both in gaseous and aqueous phase (Fig. S2). This asymmetry is due to hot bands at elevated temperatures similar to other gas-like species (e.g., $CO_2$ or $CH_4$)[6], and hydrogen bond dynamics in aqueous solution as known for the $HSO_4^-$ ion[7,8]. The $H_2S_{(aq)}$ asymmetric band may thus easily be mistaken for the $HS^-$ $\nu_1$ peak. Thus, reaction (1) is generally insignificant at shallow-crust magmatic contexts contrary to the major conclusions of ref.[1].

The related critical issue in ref.[1] is about the significance of the $[S_3^{\bullet}]^-$ ion in geological fluids[2-5]. I remind here that Pokrovski and Dubessy[2] derived the thermodynamic properties of $[S_3^{\bullet}]^-$ from the equilibrium constant ($K_2$) of the formal reaction

$$2H_2S_{(aq)} + SO_4^{2-} + H^+ = [S_3^{\bullet}]^- + 0.75O_{2(g)} + 2.5H_2O \quad (2)$$

based on Raman spectroscopy analyses in different optical cell configurations (diamond-anvil cells and capillaries) at 3 different laser wavelengths[2,3], covering 2 log units of $[S_3^{\bullet}]^-$ concentrations, along with UV-Vis absorption spectroscopy measurements[9]. These independent data are all in mutual agreement, confirming the $K_2$ value validity across the whole $T$-$P$ range covered (25–500 °C, 1 bar–15 kbar), with no evidence of potential artefacts related to Raman signal absorption in coloured solutions[10]. Reaction (2) follows standard thermodynamic formalism and in no way does it reflect any molecular mechanism of $[S_3^{\bullet}]^-$ formation as erroneously interpreted by Farsang and Zajacz[1].

The sulfur speciation scheme of Farsang and Zajacz[1] (Fig. 1a equivalent to Fig. 2c in ref.[1]) suffers from the following flaws that led to artificial overestimation of $[S_3^{\bullet}]^-$ concentrations. First, there is an inconsistency with the data source chosen for $H_2S$[12] (Table S5 in ref.[1]). In the thermodynamic framework of ref.[3], the $K_2$ values are bound to the thermodynamic properties of $H_2S_{(aq)}$ from ref.[11] consistent with the HKF equation of state. Although the difference in $H_2S_{(aq)}$ Gibbs energy values between the two sources[11,12] at ≤500 °C is within experimental uncertainties of few kJ/mol[3], it significantly increases towards magmatic temperatures (reaching 25 kJ/mol at 875 °C, 2 kbar), thereby impacting all sulfur species distribution. A correction of –2.3 log units to the $K_2$ value is required to keep up with the values used in ref.[3]. Second, the model[1] ignores the formation of alkali ion pairs with sulfur anionic species, which become increasingly abundant at elevated temperatures[3]. Third, the calculated $H_2S/SO_2$ transition at $f_{O2}$~NNO+1.5 (Fig. 1a) is inconsistent with the experimentally derived equilibrium constant ($K_3$) of the reaction below yielding $f_{O2}$~NNO+0.3[1].

$$H_2S_{(aq)} + 1.5O_{2(aq)} = SO_{2(aq)} + H_2O \quad (3)$$

Following the reaction stoichiometry, a correction of about +2 log units for $K_3$ is required in the model to be comparable with the experiment.

Fig. 1c shows calculated sulfur speciation by collectively applying the three corrections above, in line with the [$S_3^{•}$]$^-$ thermodynamic framework[3] and the experimental data[1]. The resulting [$S_3^{•}$]$^-$ is 250(!) times smaller (mole fraction ≤0.0004 vs. 0.1 in ref.[1]), undetectable by non-resonant Raman. Regretfully, Farsang and Zajacz[1] provide neither detection limit estimations nor error assessment of their Raman analyses. The signal arising from the $HSO_4^-$ moiety was not detected even at strongly oxidizing conditions (Table $S_2$ in ref.[1]) whereas hydrosulfate predicted mole fractions are ~0.01 (Fig. 1b), which could be an optimistic detection limit in the configuration of ref.[1]. Low [$S_3^{•}$]$^-$ concentrations (<0.005), unlikely detectable by non-resonant Raman, are also predicted at lower temperatures (Fig. S3 at 450 °C, 500 bar), corresponding to the analyses of re-heated fluid inclusions[1]. Thus, in contrast to the authors' own conclusions[1], their study provides no evidence that would allow one to question the validity of the thermodynamic properties of [$S_3^{•}$]$^-$, $HS^-$ and $H_2S_{(aq)}$ and their metal complexes in hydrothermal fluids.

The experimental compositions used in ref.[1] do not allow quantification of the sulfur radical ions and, therefore, cannot be used to infer its abundance in nature. This is because they were dominated by NaCl (or KCl) salts imposing $Na^+ \approx Cl^-$, with no excess of cations to allow for an electrical balance favorable for formation of negatively charged [$S_3^{•}$]$^-$ (and $HS^-$). In nature, by contrast, magmatic $H_2S/SO_2$ acidic fluids interact with alkali aluminosilicate melts or rocks that supply extra alkali and alkali-earth cations as well as increase the fluid pH. As shown in Fig. 2, bringing the fluid composition of ref.[1] which corresponds to the maximum [$S_3^{•}$]$^-$ mole fraction in Fig. 1c (which is only <0.001), into equilibrium with granitic rocks, generates [$S_3^{•}$]$^-$ fractions of ~0.1 (30 % total sulfur) between 600 and 300 °C Thus the "exotic" trisulfur ion at ref.[1] conditions becomes an important form of sulfur in porphyry-deposit contexts[3,13]. As controlled by all inter-related parameters, such as sulfur content, redox, pH, alkalinity, and cation-anion balance, radical sulfur ions and their metal complexes can be significantly favored in various magmatic-hydrothermal and metamorphic fluids as quantitatively demonstrated in recent studies[3,5,13–15].

**Acknowledgements**

This work is supported by the Mission pour les initiatives transverses et interdisciplinaires (MITI) interdisciplinary programs of the CNRS "MetalloMix 2021" (grant PtS3) and "Conditions Extremes 2024" (grant ExtremeS) and by the Institute Carnot ISIFoR (grants OrPet and AsCOCrit).

**Author Contributions**

G.S.P. is the sole author of this contribution who designed and conducted the study, made the thermodynamic analysis, and wrote the manuscript.

**Competing interests**

The author declares no competing interests.

**Supplementary Information**

All numerical data used in the thermodynamic and Raman spectra analyses and plotted in manuscript figures are supplied as an excel file. Additional data are available upon request.

## Main text figures

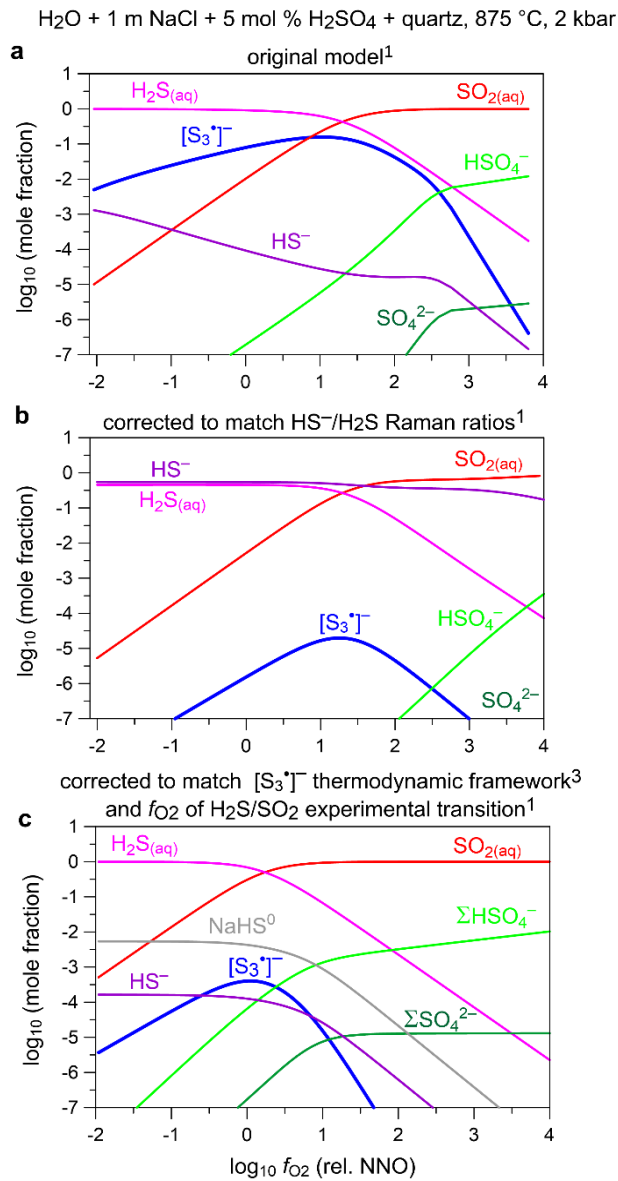

**Fig. 1. Comparison of sulfur speciation models in the aqueous fluid phase at the conditions of Farsang and Zajacz study[1].** Species concentrations are in mole fractions of total sulfur. Oxygen fugacity is relative to the nickel-nickel oxide conventional buffer (NNO), with an absolute $\log_{10}f_{O2}$ value of −12.36. **(a)** The original model (equivalent to Fig. 2a and S4 of ref.[1]) reproduced using thermodynamic data of Table S5 in ref.[1], yielding mole fractions of $[S_3^{\bullet}]^-$ up to 0.1 (equivalent to ~30% $S_{tot}$) and $HS^-$ of <0.001. **(b)** A model consistent with the reported $HS^-/H_2S$ ratios of >1 from the Raman spectra at reducing conditions[1]. To match such ratios, the reaction (1) dissociation constant was increased by 8 log units, equivalent to a correction of +176 kJ/mol for the $HS^-$ Gibbs energy from the original model **(a)**, yielding $[S_3^{\bullet}]^-$ fractions <$10^{-5}$. **(c)** A model consistent with *i)* thermodynamic framework within which the $[S_3^{\bullet}]^-$ parameters have been derived in ref.[2], with $H_2S_{(aq)}$ thermodynamic properties from ref.[9], and including those of Na ion pairs ($NaHSO_4^0$, $NaSO_4^-$ and $NaHS^0$) from ref.[3]; and *ii)* the experimental $f_{O2}$ value of NNO+0.3 for the $H_2S/SO_2$ transition[1] as per the reaction (3) equilibrium constant[4] (equivalent to a correction of −50 kJ/mol for $SO_{2(aq)}$ Gibbs energy value of ref.[9]). The resulting mole fractions of the $[S_3^{\bullet}]^-$ are ≤0.0004 (*vs* 0.1 in **(a)**), which are undetectable by non-resonant Raman, in perfect agreement with its thermodynamic properties reported by Pokrovski and Dubessy[2].

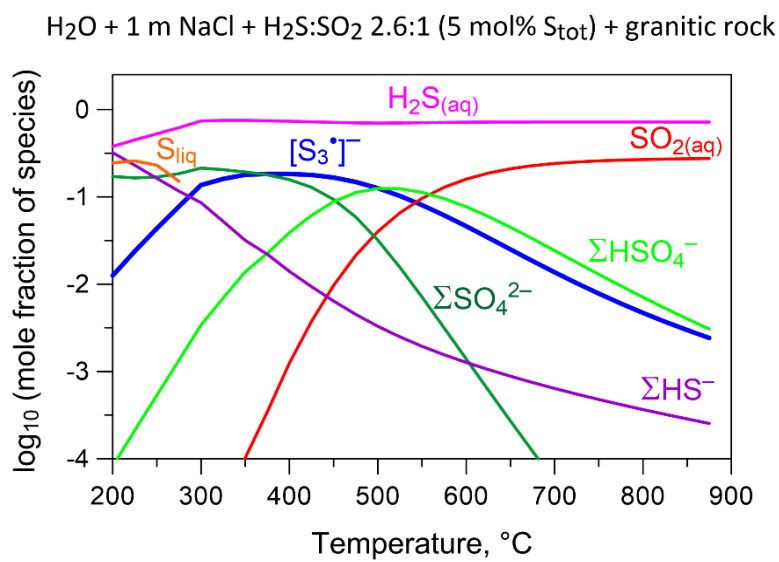

**Fig. 2. Concentrations of sulfur species (expressed in mole fraction of total sulfur) calculated in an aqueous magmatic fluid of typical composition of ref.[1] brought into equilibrium with a granitic rock.** The fluid with an $H_2S/SO_2$ molal ratio of 2.6 (corresponding to the maximum of $[S_3^{\bullet}]^-$ abundance in Fig. 1) is equilibrated with the quartz-muscovite-potassic feldspar-albite mineral assemblage at temperatures from 875 to 200 °C and pressures from 2.0 to 0.2 kbar, respectively, simulating fluid evolution upon cooling in a magmatic-hydrothermal porphyry-deposit context. Data sources for all species and minerals are from Table C.1 of ref.[1] consistent with the framework of $[S_3^{\bullet}]^-$ thermodynamic properties. The $\Sigma$ sign stands for the sum of sulfate, hydrosulfate and hydrosulfide forms (that include both anions and their ion pairs with K and Na). Concentrations of Na plus K are from 0.9 to 2.6 molal from 875 to 200 °C, and pH range is 5.5–6.2. The curve breaks at ~300 °C are due to precipitation of liquid sulfur. Remarkably, mole fractions of $[S_3^{\bullet}]^-$ being <0.001 at conditions of ref.[1] (Fig. 1b) attain 0.1–0.2 in the same fluid under conditions of porphyry deposit formation.

## Supplementary Extended Data Figures

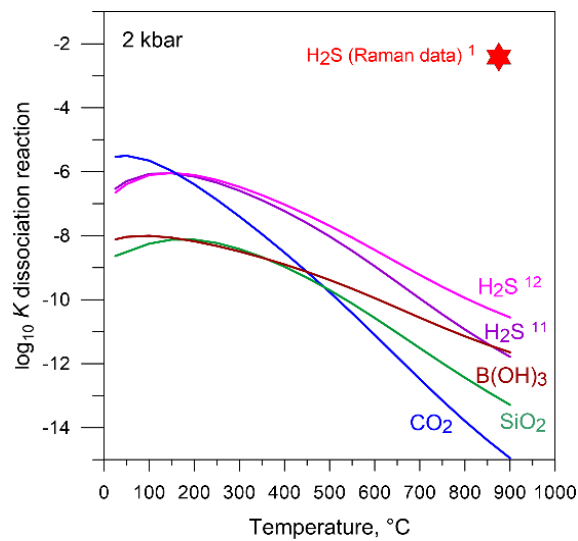

**Extended Data Fig. S1. Dissociation constants of $H_2S_{(aq)}$ (reaction 1) as a function of temperature at 2 kbar and their comparison with chemically analogous weak acids**. The curves were generated using the thermodynamic data for $H_2S$ from ref. 11 and 12 for reaction (1) and those for other species from the SUPCRT92 database, http://geopig3.la.asu.edu:8080/GEOPIG/index.html, corresponding to the following reactions: $CO_2 + H_2O = HCO_3^- + H^+$; $SiO_2 + H_2O = HSiO_3^- + H^+$; $B(OH)_3 = H_2BO_3^- + H^+$. The red star denotes the minimal $K_1$ value necessary to match the $H_2S/HS^-$ ratios of >1 reported by Farsang and Zajacz[1] from their Raman spectra analysis. Such an anomalous value is likely to be an artifact of Raman spectra assignment for $HS^-$ in ref.[1].

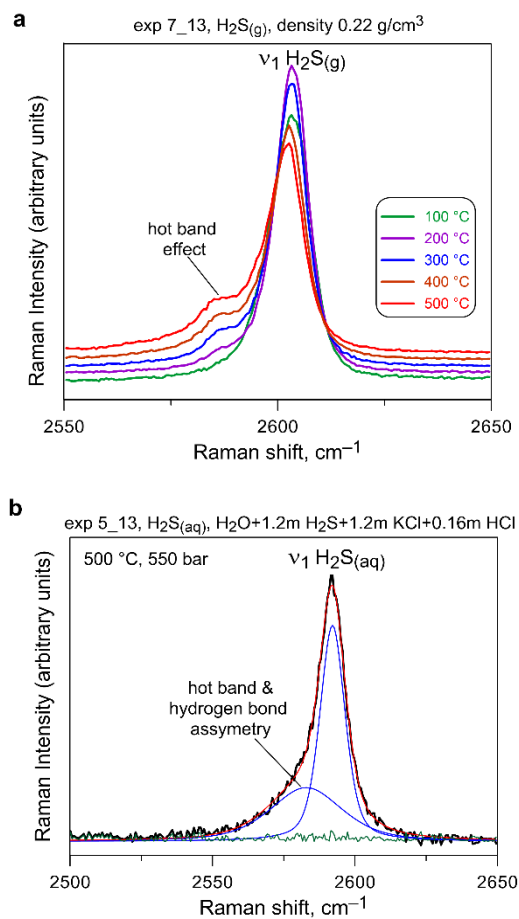

**Extended Data Fig. S2. Raman spectra of gaseous and aqueous H₂S showing the intrinsic asymmetry of the H₂S ν₁ band.** Data are from ref.[3] (their Table B.1) for the H₂S-bearing experimental fluid compositions indicated in the figure. Spectra were recorded with 457 and 514 nm lasers in a capillary cell. **(a)** Hot bands growing with temperature for H₂S$_{(gas)}$. **(b)** H₂S$_{(aq)}$ peak asymmetry in strongly acidic aqueous supercritical fluid at 500 °C and 550 bar in which the HS⁻ ion or alkali ion pair contribution is <0.01% of total H₂S. The experimental spectrum is shown in black along with its total fit (red) with 2 contributions (blue) and fit residuals (green). The lower-wavenumber band is strikingly similar to that of ref.[1] (their Fig. 1 and S15), and could easily be mistaken for the HS⁻ ν₁ peak having a very similar Raman shift position (~2575 cm⁻¹).

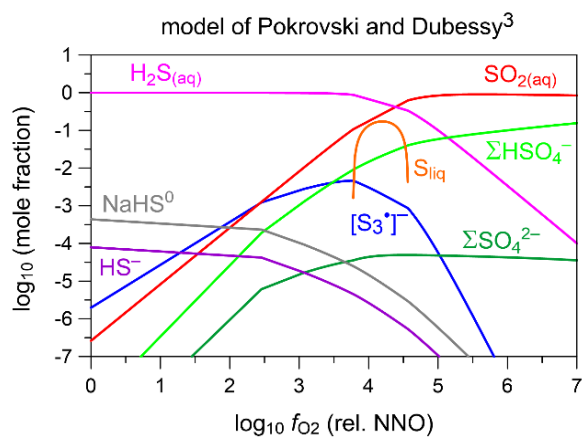

**Extended Data Fig. S3**. **Speciation of sulfur predicted at 450°C and 500 bar for a typical fluid composition of ref.[1] (indicated in the figure) as a function of oxygen fugacity (relative to the nickel-nickel oxide conventional buffer, NNO)**. These conditions match the Raman analyses in ref.[1] of re-heated fluid inclusions along a water isochore of ~0.4 g/cm³ that were synthesized at 875°C and 2 kbar. The thermodynamic data sources are from Pokrovski and Dubessy[3], consistent with the framework of the $[S_3^•]^-$ properties. Curve breaks are due to formation of liquid sulfur in a narrow range of $f_{O2}$. The absolute $\log f_{O2}$ (in bars) value of the NNO reference is −24.95. The $f_{O2}$ value of the $H_2S/SO_2$ transition is shifted to ~NNO+4 compared to the magmatic temperatures of Fig. 1, intrinsic to $f_{O2}$ change in aqueous solution on cooling[3]. Note that $[S_3^•]^-$ mole fractions are below <0.005, which would be undetectable by non-resonance Raman in perfect agreement with the ref.[1] observations, but in sharp contrast to the authors' own conclusions that the thermodynamic properties of $[S_3^•]^-$ were incorrect.